\documentclass[a4paper,12pt]{article}
\usepackage[utf8]{inputenc}
\usepackage{pifont,textcomp}
\usepackage{amsfonts}
\usepackage{amsmath}
\usepackage{amscd}
\usepackage{array}
\usepackage{multirow}
\usepackage{mathrsfs}
\usepackage{amssymb}
\usepackage{amsthm}
\newcommand{\beqn}{\begin{eqnarray}}
\newcommand{\eeqn}{\end{eqnarray}}
\newcommand{\beq}{\begin{equation}}
\newcommand{\eeq}{\end{equation}}
\newcommand{\bpro}{\begin{proposition}}
\newcommand{\epro}{\end{proposition}}
\newcommand{\blem}{\begin{lemma}}
\newcommand{\elem}{\end{lemma}}
\newcommand{\bdfn}{\begin{definition}}
\newcommand{\edfn}{\end{definition}}
\newcommand{\bcor}{\begin{corollary}}
\newcommand{\ecor}{\end{corollary}}
\newcommand{\bthm}{\begin{theorem}}
\newcommand{\ethm}{\end{theorem}}
\newcommand{\bex}{\begin{example}}
\newcommand{\eex}{\end{example}}
\newcommand{\brmq}{\begin{remark}}
\newcommand{\ermq}{\end{remark}}
\newcommand{\benum}{\begin{enumerate}}
\newcommand{\eenum}{\end{enumerate}}
\newcommand{\bitem}{\begin{itemize}}
\newcommand{\eitem}{\end{itemize}}

\title{The damped harmonic oscillator at the classical limit of the Snyder-de Sitter space}
\author{ Latévi M. Lawson$^{1}$,  Ibrahim Nonkan\'e$^{2}$ and  Komi Sodoga$^{3}$ \\\\
	 ${}^1$African Institute for Mathematical Sciences (AIMS) Ghana\\
	Summerhill Estates, East Legon Hills, Santoe, Accra\\
	P.O. Box LG DTD 20046, Legon, Accra, Ghana\\\\
	${}^2$Départément d'économie et de mathématiques appliquées,\\ IUFIC, Université Ouaga II, Burkina Faso\\\\
	${}^3$Université de Lomé, Faculté des Sciences, Departement de Physique,\\
	Laboratoire de Physique des Matériaux et des Composants\\
	à Semi-Conducteurs, 01 BP 1515 Lomé, Togo\\\\
	latevi@aims.edu.gh$^{1}$,\\
 inonkane@univ-ouaga2.bf$^{2}$,	 antoinekomisodoga@gmail.com$^{3}$
\space\\
}

\begin{document}

\maketitle

\begin{abstract} 
 Valtancoli in  his paper entitled [P. Valtancoli, {\it Canonical transformations and minimal length}, J. Math. Phys.  56, 122107 (2015)] has shown how the deformation of the  canonical transformations can be made compatible  with the deformed 
 Poisson brackets. Based on this  work and through an appropriate canonical transformation,  we solve the problem of  one dimensional (1D) damped harmonic oscillator at the classical limit  of the  Snyder-de Sitter (SdS) space. We show that the  equations of  the motion can be described by trigonometric functions with frequency and period depending on the deformed  and the damped parameters. We eventually  discuss about the influences of these parameters on the motion of the system. 
 \space 

\end{abstract}

\section{Introduction}

The search  of quantum gravity is one the active field of research that has attracted much attention in the last decades. Among all the candidate  theories \cite{2,3,4} to address  this problem, the non-commutative geometry seems to be the promising  approach to quantum gravity. In this sense, more recently, Lawson has  proposed  a  model of   quantum non-commutative geometry  \cite{6,6'} that may  describe the space-time at Planck scale. The intersting physical result obtained  in this theory which differs from the similar theories \cite{7,8} is that, this non-commutative space-time leads to minimal and maximal lengths of graviton for simultaneous measurement. The existence of this maximal length which is the basic difference to the one of the minimal length scenario, brings a lot of new features in the representation of this  space and could be the approach candidate to the measurement of quantum gravity with energies currently accessible in a laboratory.

However, more general models of non-commutative spaces  exist and describe the structure of  spacetime at short  distance and preserve the Lorentz invariance. The best known is the Snyder model \cite{9}, which is the first  attempt of  introducing  a fundamental length scale and is invariant under the Lorentz group. This model is generalized to a spacetime background of constant curvature namely the Snyder-de Sitter (SdS) model \cite{10}. It is an example of non-commutative spacetime admitting three fundamental parameters and  is also called Triply Special Relativity (TSR) and is invariant under the action of the de Sitter group \cite{10}. Some applications of this work have been done in \cite{11} and the model of  a free particle and of an oscillator   have been solved  in this framework.

A characteristic of non-commutative spaces is that the corresponding classical phase space is not canonical i.e. the Poisson brackets do not have the
usual form. Thus, at the classical limit of the SdS space  the solutions of  free particle and  harmonic oscillator  systems had been  also obtained 
 by substituting the generalized commutations  with the deformed Poisson brackets \cite{12}. Hence, in this paper we are interesting in the  study of a one dimensional ($1$D) damped  harmonic oscillator \cite{13,14} in the deformed Poisson brackets.  Since this system is explicitly time-dependent, to determine its solutions of motion,   we first make the Hamiltonian simpler by means of a suitable canonical transformation. Then we solve the equations of motion for the   new canonical variables in the deformed Poisson brackets. We provide  the solutions of the motion  in terms of trigonometric functions   where the  frequency  and  the period of   the motion depend on the  deformed  and the damped parameters. We show  that when the   damped parameter is less than a certain value of the deformed parameter, the gravity induces faster  motion of system but when it is greater than  this value the dissipation slows down the motion of the system.

   This paper is organized  as follows. In section (\ref{section2}), we review some properties of  quantum  and classical limit of SdS model in 1D. We extend in section (\ref{section3}), the classical  procedure on the  canonical transformation.
  Based on this study, we explicitly solve  in section (\ref{section4})  the 1D damped harmonic oscillator in the classical limit of SdS model.  We provide  the solutions of the motion  in terms of trigonometric functions where we discuss the influence of the deformed and the friction parameters on the motion of the system. We present our conclusion in section (\ref{section5}).

 .

\section{Classical  Snyder-de Sitter space}\label{section2}

The non relativistic quantum SdS model i.e  the Snyder model restricted to a three-dimensional sphere is generated  by  the positions operators
$\hat x_\mu$, the momentum operators $\hat p_\nu$ and the Lorentz
generators $\hat J_{\mu\nu}$  such as
\begin{eqnarray}\label{p5}
 [ \hat x_\mu,\hat x_\nu]&=&i\beta^2 \hat J_{\mu\nu},\quad  [ \hat p_\mu,\hat p_\nu]=i\alpha^2 \hat J_{\mu\nu}, \quad \mu,\nu=0,...3\cr
  [ \hat x_\mu,\hat p_\nu]&=&i\left(\eta_{\mu\nu}+\alpha^2\hat x_\mu\hat x_\nu+\beta^2 \hat p_\mu \hat p_\nu+\alpha\beta(\hat x_\mu \hat p_\nu+\hat p_\mu\hat x_\nu-\hat J_{\mu\nu})\right).
 \end{eqnarray} 
where  $\eta=diag(-1,1,1,1)$ is the flat metric and the  coupling constants  $\beta,\alpha$ ($\alpha\beta\ll 1$) have dimension of inverse length and inverse mass, respectively. They are usually identified with the square root of the cosmological constant $\alpha=10^{-24}cm^{-1}$ and with the inverse of the Planck mass, $\beta=10^5g^{-1}$ \cite{11}. The  Lorentz
generator with their standard action
on the position and momentum operators $\hat x_\mu$ nd $\hat p_\nu$ satisfy the usual commutation relations such as 
\begin{eqnarray}
[\hat J_{\mu\nu},\hat p_\mu]&=&i(\eta_{\nu\lambda}\hat p_\mu-\eta_{\nu\lambda}\hat p_\nu),\quad [\hat J_{\mu\nu},\hat p_\mu]=i(\eta_{\nu\lambda}\hat x_\mu-\eta_{\mu\lambda}\hat x_\nu),\cr
[\hat J_{\mu\nu},\hat J_{\rho\sigma}]&=& i(\eta_{\nu\rho}\hat J_{\nu\sigma}-\eta_{\sigma\mu}\hat J_{\rho\nu}-\eta_{\sigma\nu}\hat J_{\rho\mu}),\quad\quad \hbar=1.
\end{eqnarray}
The limit $\alpha\rightarrow 0$  the SdS space (\ref{p5}) gives the flats Snyder space \cite{9}
\begin{eqnarray}\label{p5}
[ \hat x_\mu,\hat x_\nu]=i\beta^2 \hat J_{\mu\nu},\quad  [ \hat p_\mu,\hat p_\nu]=0, \quad 
[ \hat x_\mu,\hat p_\nu]=i\left(\eta_{\mu\nu}+\beta^2 \hat p_\mu \hat p_\nu \right),
\end{eqnarray}
 while the limit $\beta\rightarrow 0$ yields the Heisenberg algebra of quantum mechanics in a de Sitter background endowed with projective coordinates \cite{12}.
 
 In one-dimensional case, the algebra  (\ref{p5}) is reduced into
\begin{eqnarray}\label{p6}
[ \hat x,\hat p]=i\left(1+\alpha^2\hat x^2+\beta^2 \hat p^2+\alpha\beta(\hat x \hat p+\hat p\hat x)\right),\quad [ \hat x,\hat x]= 0=[ \hat p,\hat p].
\end{eqnarray} 
and  for the simple case $\langle \hat x\rangle=0=\langle \hat p\rangle$, the uncertainty relation is given by
\begin{eqnarray}
\Delta x\Delta p\geq\frac{1}{2}\frac{\alpha^2 \Delta x+\beta^2 \Delta p }{1+\beta\alpha}.
\end{eqnarray}
If $\alpha, \beta > 0$, they imply the existence of both minimal position and momentum uncertainties, given by
\begin{eqnarray}
\Delta x=\frac{\beta}{\sqrt{1+2\alpha\beta}}=\beta(1-\alpha\beta),\quad 
\Delta p=\frac{\alpha}{\sqrt{1+2\alpha\beta}}= \alpha(1-\alpha\beta).
\end{eqnarray}
For $\alpha, \beta < 0$, no minimal uncertainties emerge \cite{12}.

In one  dimensional  classical limit, the  commutator  (\ref{p6}) is  replaced by the deformed Poisson bracket 
\begin{eqnarray}
\{  x,  p\}=1+\left(\alpha x+\beta p \right)^2,\quad  \{ x, x\}=0= \{ p, p\}
\end{eqnarray}
The equations of motion governed  by the classical Hamiltonian $H(x,p)$ are given by
\begin{eqnarray}\label{p8}
 \dot{x}&=& \{x,H\}= \frac{\partial H}{\partial p}\{  x,  p\}=¨\frac{\partial H}{\partial p}\left(1+\left(\alpha  x+\beta  p\right)^2\right),\\
 \dot{p}&=& \{p,H\}= -\frac{\partial H}{\partial x}\{  x,  p\}=¨-\frac{\partial H}{\partial x}\left(1+\left(\alpha  x+\beta  p\right)^2\right).
\end{eqnarray} 
\section{Deformed canonical transformation}\label{section3}
It is well known  in the formulation of classical mechanics that, the  transition from the canonical variables $x$ and $p$ to new arbitrary 
canonical variables  $X$ and $P$ which is called canonical transformation lets the physics of the system invariant. Therefore, the canonical transformation of the variables  $x$ and $p$  into  the variables $X$ and $P$ defines a new Hamiltonian
$K(X,P)$. It is defined by the map 
\begin{equation}  \label{p10}
  x, p, H(p,x)\longmapsto X,P,K(X,P),
\end{equation}
and the fundamental Poisson brackets read
\begin{eqnarray} \label{p11}
 \{X,P\}_{X,P}&=& \{X,P\}_{x, p}=\{x,p\}_{X,P}=1,\\
 \{X,X\}_{X,P}&=&\{P,P\}_{X,P}=0.
\end{eqnarray}
The equations of motion are given by
\begin{eqnarray}  \label{p12}
  \dot{X}=\{X,K\},\,\,\,\,\, \dot{P}=\{P,K\}.
\end{eqnarray}

Moreover, under this transformation the old Hamiltonian $H(P,X)$ is  transformed  into the new Hamiltonian as follows 
\begin{equation} \label{p14}
 K=H+\frac{\partial F}{\partial t},
\end{equation}
where  $F$ is the generating function. 
Now, if we are convinced that the Poisson brackets are invariant under the canonical transformation, then the deformation of the Poisson 
brackets must  be also invariant under this transformation \cite{1}, i.e
\begin{equation}  \label{p15}
 \{x,p\}=1+\left(\alpha +\beta p\right)^2\longrightarrow \{X,P\}=1+\left(\alpha X+\beta P\right)^2.
\end{equation}

The time evolution of the coordinate $X$ and the momentum $P$ is given by
\begin{eqnarray} \label{p17}
  \dot{X}&=&\{Q,K\}=\frac{\partial K}{\partial P}\left(1+\left(\alpha X+\beta P\right)^2\right),\\
  \dot{P}&=&\{P,K\}=-\frac{\partial K}{\partial X}\left(1+\left(\alpha X+\beta P\right)^2\right).
\end{eqnarray}
We are now in position to apply all of these aspects on the damped harmonic oscillator in order  to study  its equations of motion. 
\section{Damped harmonic oscillator in SdS space} \label{section4}
The   damped harmonic oscillator is one of the most fascinating systems  that
have remained over the years a constant source of  inspiration in  quantum physics \cite{13,14}. 
 It has attracted much attention in the literature \cite{15,16,17,18,19,20,20',20''} since the  problem related to this system 
 is far from having a satisfactory solution. In fact the  quantization of  dissipative  systems    well known in the  literature as  Caldirola and Kanai  system \cite{13,14}  has been criticized for violating certain laws of quantum theory. 
  Recently, we provide a simple and complete solution \cite{21}  to this problem using the Lewis-Riesenfeld 
 procedure \cite{22}. Therefore, in the present situation we are interested to the classical motion of this system in the minimal length scenario.
 In one dimension, its Hamiltonian   is given by
\begin{equation} \label{p18}
 H(p,x)=e^{-\gamma t}\frac{p^2}{2}+\frac{\omega_0^2}{2}x^2e^{\gamma t},
\end{equation}
where we set  $m=1$, $\gamma$ is the constant coefficient of friction and $\omega_0$ is the time-independent harmonic frequency. Since this Hamiltonian is explicitly time-dependent, it does not represent a conserved quantity.
So, we achieve its energy conservation through the time-dependent canonical transformation given  by the 
generating function of the second kind \cite{23}
\begin{equation} \label{p19}
 F(x,p,t)=xpe^{\frac{\gamma t}{2}}-\frac{\gamma}{4}x^2e^{\gamma t}.
\end{equation}
Perfoming  the following  transformation equations
\begin{eqnarray} \label{p20}
  X=\frac{\partial F}{\partial p},\,\
  P=\frac{\partial F}{\partial x},
\end{eqnarray}
we obtain the new canonical variables
\begin{eqnarray} \label{p21}
 X=xe^{\frac{\gamma}{2}t},\quad
 P=pe^{-\frac{\gamma t}{2}}+\frac{\gamma}{2}xe^{\frac{\gamma t}{2}}.
\end{eqnarray}
The relations between old $(p,x)$ and new $(P,X)$ coordinates can be written as
\begin{eqnarray} \label{p22}
 \binom{p}{x}=\begin{pmatrix} e^{\frac{\gamma}{2} t} & -\frac{\gamma}{2} e^{\gamma t} \\ 0 & e^{-\frac{\gamma}{2} t} \end{pmatrix} \binom{P}{X},\quad
 \binom{P}{X}=\begin{pmatrix} e^{-\frac{\gamma}{2} t} & \gamma e^{\frac{\gamma}{2} t} \\ 0 & e^{\frac{\gamma}{2} t} \end{pmatrix} \binom{p}{x}.
\end{eqnarray}
Their Poisson brakets are  canonical
\begin{equation}
 \{X,P\}=1=\{x,p\}.
\end{equation}
The transformed Hamiltonian is given by 
\begin{equation}
 K(X,P,t)=\frac{P^2}{2}+\frac{\omega^2}{2}X^2,
\end{equation}
where 
\begin{equation}
 \omega^2=\omega_0^2-\frac{\gamma^2}{4},
\end{equation}
is the modified frequency. Since the  new Hamiltonian is time independent, its  conserved energy is 
\begin{equation}\label{p27}
 E= \frac{P^2}{2}+\frac{\omega^2}{2}X^2.
\end{equation}
The deformed Poisson brackets  between the transformed canonical variables are
\begin{eqnarray}\label{p22}
 \{X,P\}=1+\left(\alpha X+\beta P\right)^2,\quad
 \{X,X\}=0=\{P,P\}.
\end{eqnarray}
The equation of motion reads as 
\begin{eqnarray}
 \dot{X}&=&\left( 1+\left(\alpha X+\beta P\right)^2\right)P,\label{p23}\\
 \dot{P}&=&-\omega^2 \left( 1+\left(\alpha X+\beta P\right)^2\right)X.\label{p24}
\end{eqnarray}
After some computations, the solutions of the above equations  are given by 
\begin{eqnarray}
 X&=& A(t)\left(\frac{\alpha}{\omega}\sin\varpi t-\beta\sqrt{1+2\kappa E}\cos \varpi t\right)\\
 P&=&B(t)\left( \beta \sin\varpi t+\frac{\alpha}{\omega}\sqrt{1+2\kappa E}\cos\varpi t \right)
\end{eqnarray}
where  
\begin{eqnarray}
\kappa&=&\beta^2+\frac{\alpha^2}{\omega^2},\quad \varpi=\omega\sqrt{1+2\kappa E},\cr
\omega A(t)&=& B(t)=\sqrt{\frac{2E}{\kappa(1+2\kappa E\cos^2 \varpi t)}}.
\end{eqnarray}
The solutions are periodic, but not sinusoidal, and the frequency now depends on the energy of the oscillator. In the limit $\alpha\rightarrow 0$, one
recovers the flat Snyder oscillator \cite{9}.

Returning to the original variables $x(t)$ and $p(t)$ of  the Hamiltonian (\ref{p18}) and  with the help of the relation (\ref{p21})
 we have:
\begin{eqnarray}
x(t)&=& C(t)\left(\frac{\alpha \sin\Omega t}{\sqrt{\omega_0^2-\gamma^2/4}} -\beta\sqrt{1+2\kappa E(t)}\cos\Omega t \right)\\
p(t)&=&D(t)\left( \beta \sin\Omega t+\frac{\alpha \sqrt{1+2\kappa E(t)}\cos\Omega t}{\sqrt{\omega_0^2-\gamma^2/4}} \right)\cr&&
-G(t) \left(\frac{\alpha \sin\Omega t}{\sqrt{\omega_0^2-\gamma^2/4}} -\beta\sqrt{1+2\kappa E(t)}\cos\Omega t \right)
\end{eqnarray}
where
\begin{eqnarray}
\Omega(t)&=&\sqrt{(\omega_0^2-\gamma^2/4)(1+2\kappa E(t))}\\
 C(t)&=&e^{-\frac{\gamma t}{2}}A(t)= \frac{e^{-\frac{\gamma t}{2}}}{(\omega_0^2-\gamma^2/4)}\sqrt{\frac{2E(t)}{\kappa(1+2\kappa E(t)\cos^2 \Omega t)}}\\
 D(t)&=& e^{\frac{\gamma t}{2}}B(t)= e^{\frac{\gamma t}{2}} \sqrt{\frac{2E(t)}{\kappa(1+2\kappa E(t)\cos^2 \Omega t)}}\\
 G(t)&=& \frac{\gamma e^{\gamma t}}{2} C(t).
\end{eqnarray}

In   first-order of the deformed parameter  $\kappa$ and  the damped parameter $\gamma$,  the frequency $\Omega$ and the period
$T$ of the motion are given by
\begin{eqnarray}
 \Omega&=&\omega_0\left(1-\frac{\gamma^2}{8\omega_0^2}+\kappa E(t)\right),\\
 T&=&\frac{2\pi}{\omega_0}\left(1+\frac{\gamma^2}{8\omega_0^2}-\kappa E(t)\right),
\end{eqnarray}
where we neglected the term of order $\beta \gamma^2$. 
 To zero-order in $\kappa$ and $\gamma$ i.e  for $\kappa=0=\gamma$, we recover the ordinary solutions of harmonic oscillator with frequency 
 $\omega_0$ and period $T_0=\frac{2\pi}{\omega_0}$. 
 If $\gamma<2\omega_0\sqrt{2\kappa E(t)}$, the period of
 the motion is shorter than the ordinary one $T<T_0$, i.e  the deformation of 
 Poisson bracket  which is the manifestation of an effect of gravity induces a faster motion of the oscillator.
 But for $\gamma>2\omega_0\sqrt{2\kappa E(t)}$, the period of the motion increase which means that the dissipation of the system slows down the 
 motion of the oscillator.
\section{Conclusion}\label{section5}
In this article, we first studied the properties of  the $1$D deformed Poisson brackets which  are regarded as the classical limit of the
generalized Heisenberg commutators.  Then, we  reviewed the innovative  concept of $\beta$-canonical transformation introduced by 
Valtancoli \cite{1}  appointed in this work as the deformed canonical transformation. This concept allowed us to maintain the invariance 
of the deformation of the Poisson brackets in the reparametrization of the phase space variables. We applied all these theories to
 the well-known one dimensional damped harmonic oscillator. With an appropriate canonical transformation, we transformed the time-dependent  Hamiltonian into the time-independent Hamiltonian.
 Based on  \cite{1}   we have shown that the solutions  of the equations of motion  can  be expressed in terms
 of trigonometric functions  with the 
 frequency and period depending on the deformed and the damped parameters of the system. Finally,  we  discussed about the influences these
 parameters on the motion. We have shown that when the deformed parameter is greater than the damped parameter
 the period of the motion is shorter, conversely the period increases when the damped parameter is greater than the deformed parameter.

\section*{Acknowledgments}
L.M Lawson acknowledges support from AIMS-Ghana
under the Postdoctoral fellow/teaching assistance (Tutor) grant

\end{document}